\documentclass[prl,twocolumn,superscriptaddress]{revtex4}
\usepackage{graphicx}
\usepackage{amsfonts}
\usepackage{amsmath}
\usepackage{amssymb}
\usepackage{bm}
\usepackage{slashed}
\usepackage{dsfont} 

\usepackage{color}
\usepackage{soul}

\begin{document}
\title{Kondo signatures in Dirac spin liquids: \\
Non-Abelian bosonization after Chern-Simons fermionization}
\author{Rui Wang}
\affiliation{National Laboratory of Solid State Microstructures and Department of Physics, Nanjing University, Nanjing 210093, China}
\affiliation{Collaborative Innovation Center for Advanced Microstructures, Nanjing 210093, China}
\author{Yilin  Wang}
\affiliation{Department of Condensed Matter Physics and Materials Science,
Brookhaven National Laboratory, Upton, New York 11973, USA}
\author {Y. X. Zhao} 
\affiliation{National Laboratory of Solid State Microstructures and Department of Physics, Nanjing University, Nanjing 210093, China}
\affiliation{Collaborative Innovation Center for Advanced Microstructures, Nanjing 210093, China}
\author{Baigeng Wang}
\email{bgwang@nju.edu.cn}
\affiliation{National Laboratory of Solid State Microstructures and Department of Physics, Nanjing University, Nanjing 210093, China}
\affiliation{Collaborative Innovation Center for Advanced Microstructures, Nanjing 210093, China}

\begin{abstract}
Quantum impurities serve as in-situ probes of the frustrated quantum magnets, and Dirac spin liquids are an important class of quantum spin liquids. Here, we present a general method,  a combination of the Chern-Simons fermionization and the Wess-Zumino-Witten theory, to study the quantum impurity in Dirac spin liquids.  Under the Chern-Simons fermionization, the gauge fluctuations are apparently suppressed and the low-energy physics is described by a number of Dirac valleys with valley-dependent pseudospin-momentum locking.  The ($2+1$)D effective theory can be further reduced into the ($1+1$)D Wess-Zumino-Witten theory by rotational symmetry, where the  pseudospin-exchange between Dirac fermions and the impurity can then be solved by the non-Abelian bosonization. Consequently, fixed points of Fermi liquid and non-Fermi liquid are identified, respectively, depending on the relevance of the impurity scattering among the Dirac valleys. This leads to experimental fingerprints for Dirac spin liquids,
including a Kondo-induced magneto-thermal effect, a non-monotonous thermal conductivity during the crossover, and an anisotropic spin correlation function.
\end{abstract}

\maketitle
\emph{\color{blue}{Introduction.--}} Quantum spin liquids (QSLs) \cite{pandersona,lsavary,yizhou}, strongly entangled quantum states that evade ordering down to zero temperature, pose a great challenge for their experimental observation. The main difficulty is due to the fact that the low-lying excitations of QSLs are fractionalized particles \cite{lsavary,yizhou}, whose nonlocal nature is beyond the capabilities of usual experimental probes. For a number of candidate materials, the absence of ordering is evidenced by the specific heat and the muon spin relaxation experiments at ultra-low-temperature \cite{shimizu,kanoadaa,kkanoda,lefebvre,fkagawa,shimizub,pratt,syamashita,Manna,Mendels,Helton,hanth,ocmt,singh,myamashita,Sandilands,ysli,yslib,paddison,yshenn,Nasu,Banerjeey,kran,jzhengk,shbaekk}. However, the identification of the highly-entangled liquid states remains elusive \cite{myamashita,myamashitab,yjyu,jmni,Jmnib}: the observed thermal conductivity at low temperatures suggests a dominant role played by the phonons,  obscuring the contributions from the fractionalized particles \cite{jmni,Jmnib}, e.g., the spinons, if any. Therefore, to further validate the QSL ground states, it is urgent to predict more finger-print experimental features that are unique to the fractionalized excitations \cite{lsavary}.

One strategy is to use quantum impurities as in-situ probes ~\cite{JK,wilson,ylwang,hshiba,Balatsky,huihu,ruikondo}, which can induce many-body resonance and result in global change of the thermal dynamical properties of the bath. This impurity approach has drawn growing attention and was applied to various QSLs since the last decade \cite{Khaliullin,Kolezhuk}, including the Kitaev spin liquids \cite{Vojta,kusum,sddas,Sreejith}, the spin liquids with spinon Fermi surfaces \cite{pribeiro,rldoretto}, and also the deconfined quantum critical point (DQCP) in frustrated magnets \cite{kaul,Kaj,maxa,maxab}. The Kondo signature in spinon bath was also observed by recent experiments on Zn-brochantite \cite{mgomi,mogomil}. 
However, some key questions are yet to be addressed. First, the fractionalized excitations in QSLs are usually coupled to emergent gauge fields \cite{Halperin,pribeiro,Kolezhuk}, but it is still not clear whether the gauge fields would participate in the local many-body resonance triggered by the impurity \cite{pribeiro,Serbyn}. Second, the conventional parton mean-field theory for QSLs is subject to the single-occupation condition \cite{rmplee}, and therefore inevitably requires approximations such as the large-N treatment. This is not sufficiently satisfactory.

In this Letter, we focus on quantum impurity in Dirac QSLs, and address the above questions by developing a more rigorous and systematic approach, namely the lattice Chern-Simons fermionization plus 
Wess-Zumino-Witten (WZW) theory. Based on this approach, Kondo-induced signatures of the Dirac QSLs are found.  The Dirac QSLs, whose fractionalized excitations enjoy linear dispersion, are of particular importance as they are closely related to the quantum antiferromagnetism as well as the DQCP \cite{Akihiro}.  They are also proposed as  stable ground states of  certain quantum spin models \cite{yingrana,zxliu,shijiehu}. 
We fermionize the quantum magnets using the recently developed lattice Chern-Simons (CS) fermion representation~\cite{ttased,ruia,ruib,tsedrakyana,tsedrak,ruic}, which naturally suppresses gauge fluctuations in low energies. Consequently, the low-energy physics generally corresponds to a number of Dirac valleys with valley-dependent pseudospin-momentum locking (PSML), where Dirac fermions experience the pseudospin exchange with the effective magnetic impurity. The low-energy effective model, owing to its rotation symmetry, can be reduced to ($1+1$)D conformal field theories (CFTs), namely WZW theories. 
Then, based on the non-Abelian bosonization, two types of Kondo fixed points, either the Fermi liquid (FL) or the non-Fermi liquid (NFL), are identified, depending on the relevance of the impurity scattering among the Dirac valleys. Remarkably, we show that, although both of the two fixed points are charge-insulating, they display distinct scaling behaviors in field-modulated thermal conductivities. This can serve as a  Kondo-induced characterization of the Dirac QSLs. 

\textit{\color{blue}{A general reduction to CFT.--}} Let us begin with considering the general Kondo problem with a bath of Dirac fermions: $2$D Dirac valleys labeled by $a$  related by some point group \cite{footnote2} in the Brillouin zone, whose low-energy excitations are described by
\begin{equation}\label{Dirac-fermions}
H^D=\sum_{a}\int\frac{d^2k}{(2\pi)^2}~f^{a\dagger}(\bm{k})(v_F\bm{k}\cdot\bm{\tau}^{(a)}-\mu)\,f^{a}(\bm{k}).
\end{equation}
Here,  we allow a valley-dependent PSML, such that for each valley $a$ ($a=1,2,...,k$), the set of pseudospin $\tau^{(a)}_i$ with $i=1,2$ can be different and are not necessarily the standard Pauli matrices. Since they satisfy the Clifford algebra, $\{\tau^{(a)}_i,\tau^{(a)}_j\}=2\delta_{ij}\mathds{1}_2$, we can always appropriately choose $\tau^{(a)}_3$, so that there is a unitary transformation $U^{a}$ that transforms $\tau^{(a)}_{i}$ into the standard Pauli matrices $\tau_i=U^a\tau^{(a)}_iU^{a\dagger}$. We further consider an impurity effectively characterized by $\bm{S}_{imp}$ pinned at $\bm{r}=0$ in real space.
These Dirac fermions have an effective Kondo exchange coupling with the impurity $\bm{S}_{imp}$,
\begin{equation}\label{Kondo exchange}
H'=\sum_{a,b}\lambda_{ab}~f^{a\dagger}(0)\frac{\bm{\tau}}{2} f^b(0)\cdot \bm{S}_{imp},
\end{equation}
where $\bm{\tau}=(\tau_1,\tau_2,\tau_3)$ and $\lambda$ is a symmetric real matrix. The point group symmetry that relates the Dirac valleys imposes constraints on $\lambda$. For instance, diagonal entries are all equal.
As we observe, for the Hamiltonian in Eq.\eqref{Dirac-fermions}, the density of states (DOS) vanish at $\mu=0$. At this critical point, the Kondo exchange coupling in Eq.\eqref{Kondo exchange} is irrelevant. Away from the critical point with $\mu\ne 0$, the low-energy degrees of freedom are described by the soft fermionic modes in the vicinity of the Fermi circles (see Fig.1(a)). 

\begin{figure}[t]
\includegraphics[width=\linewidth]{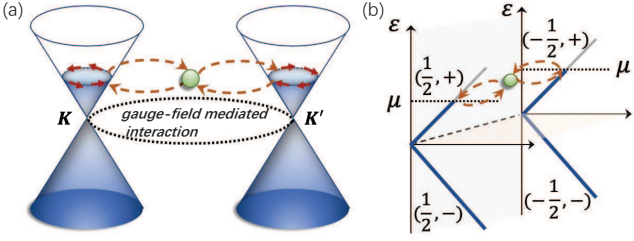}\label{fig1}
\caption{(a) The impurity is coupled to different Dirac valleys, whose Fermi levels are modulated by external magnetic field (see below).
(b) The low-energy modes of Dirac valleys correspond to chiral fermions with orbital angular momentum and pseudospin indices $(j,s)$. The impurity is only coupled to the soft modes with $s=+$ ($s=-$) and $j=\pm1/2$ for a positive (negative) chemical potential. }
\end{figure}

For technical convenience, let us transform the pseudo-spin $\bm{\tau}^{(a)}$ for each valley $a$ into the standard Pauli matrices. That is, we introduce $\tilde{f}^a(\bm{k})=U^{a\dagger}f^a(\bm{k})$.
In accord with the Kondo exchange Eq.\eqref{Kondo exchange}, it is convenient to work with the polar coordinates. 
Since the PSML of \eqref{Dirac-fermions} preserves the total angular momentum $J_z=L_z+\frac{1}{2}\tau_z$, we expand the fields as, $\tilde{f}^a(\bm{k})=\sum_{j,s}f_{j,s}^a(k)\chi_{j,s}(\phi)$, where $\chi_{j,s}(\phi)$ is the eigenstates  of $J_z$, with quantum numbers $(j,s)$, where $j$ is the half-integer eigenvalues of $J_z$ and $s=\pm$ labels the eigenvalues of $\tau_z$. In such basis, in-coming and out-going radial waves emerge with positive and negative energies, respectively (see Fig.1(b)). Then, for $\mu>0$, the soft modes are right-handed fermions described by
\begin{equation}\label{chiral-fermions}
H^D_{eff}=v_F\sum_{a,j} \int_{-\Lambda}^{\Lambda}\frac{dq}{2\pi}~\psi^{a\dagger}_{j}(q)q\psi^{a}_j(q),
\end{equation}
where $\psi_j^a(q)=u^{a}_j\sqrt{\frac{k_F}{2\pi}}\tilde{f}_{j,+}(k_F+q)$ with the index $s=+$ omitted. 
Here, an arbitrary $U(1)$ factor freedom $u^a_j$ is allowed for each soft mode $(a,j)$ for later usage. The Fermi momentum is $k_F=\mu/v_F$, and $q$ takes value within a cutoff, $q\in[-\Lambda,\Lambda]$. For $\mu<0$, analogously the soft modes corresponds to left-handed fermions, and can be treated in parallel to the case of $\mu>0$. 
Note that the renormalized soft fermion fields by $\sqrt{\frac{k_F}{2\pi}}$ satisfy the anti-commutation relations of $1$D fermions, for instance, $\{\psi^\dagger_j(q),\psi_{j'}(q')\}=2\pi\delta(q-q')$.

The Kondo exchange $H^{\prime}$ can be transformed accordingly, reducing to the coupling to $1$D soft fermions. Owing to the $U(1)$ factor $u^a_j$ of the soft modes, it can be cast into a simple form where only $j=\pm 1/2$ are relevant: 
\begin{equation}\label{Kondo_soft-modes}
H'_{eff}=\sum_{a,b}g_{ab}~\psi^{a\dagger}(0) \frac{\bm{\sigma}}{2} \psi^{b}(0)\cdot \bm{S}_{imp},
\end{equation}
where $\psi^{a\dagger}(x)=[\psi^{a\dagger}_{\frac{1}{2}}(x),\psi^{a\dagger}_{-\frac{1}{2}}(x)]$, $g_{ab}=\pi k_F\lambda_{ab}$, and $\boldsymbol{\sigma}$ denotes the Pauli matrix defined in the angular momentum space. Eq.\eqref{Kondo_soft-modes} implies a cutoff, with summation over only $j=\pm 1/2$ in Eq.\eqref{chiral-fermions}. 

The above derivation shows that general Kondo problems in 2D Dirac systems with valley-dependent PSML can be reduced to  $k$ valleys of soft fermions coupled to the impurity, which generally allows for a CFT description of the underlying infared fixed points \cite{Affleckb,Afflecka,Affleckc,Ludwig,Affleckd,Afflecke,Ludwigf,edwitten}. The single valley case is illustrated by Fig.1(b).

Let us firstly consider the case with ignorable intervalley scatterings, namely $g_{ab}=g\delta_{ab}$. Then, the Kondo exchange Eq. \eqref{Kondo_soft-modes} becomes $H'_{eff}=g\bm{J}(0)\cdot \bm{S}_{imp}$, where the $SU(2)$ current $\bm{J}(x)=\sum_a\psi^\dagger_a(x) \bm{\sigma}\psi_a(x)/2$. This motivates us to consider the global symmetry $U(1)\times SU(k)\times SU(2)$ of Eq.\eqref{chiral-fermions} for the charge, valley and pseudospin sector, separately, leading to the bosonization for the Hamiltonian density as \cite{edwitten},
\begin{equation}\label{bosonization}
\mathcal{H}^D_{eff}=\frac{\pi v_F}{2k} J^2+\frac{2\pi v_F}{k+2}\bm{J}^2+\frac{2\pi v_F}{k+2}\mathcal{J}^2,
\end{equation}
 where the currents for $U(1)$ and $SU(k)$ sector respectively read as, $J=\sum_{a,j}\psi^{a\dagger}_{j}\psi^a_{j}$ and $\mathcal{J}^A=\sum_j \psi^\dagger_j T^A \psi_{j}$, with $T^A$ the generators of $SU(k)$. Since the impurity only interacts with the pseudospin current operator $\bm{J}$, which satisfies the Kac-Moody algebras $\mathrm{SU}(2)_k$, the fermion bath can enjoy a NFL fixed point \cite{Affleckb,Afflecka} characterized by $\mathrm{U}(1)\times \mathrm{SU}(2)_k\times \mathrm{SU}(k)_2$ CFT. 
In the generic case with off-diagonal entries of $g$, the Kondo exchange Eq.\eqref{Kondo_soft-modes} violates the rotational symmetry in the valley space. Therefore, in general $\mathcal{J}$ is no longer a conserved current, and the level of $\mathrm{SU}(2)_k$ will be reduced, resulting in a different fixed point.

\textit{\color{blue}{Chern-Simons Dirac fermions in spin liquids.}}-- 
We now demonstrate how the above formalism can be related to a general 2D Dirac QSL. 
Our scheme is to utilize the CS fermionization~\cite{ttased,ruia,ruib,tsedrakyana,tsedrak,ruic} to describe the Dirac QSLs. We represent the local spin-$1/2$ state as a spinless fermion state attached with a unit of $U(1)$ gauge flux to preserve the bosonic statistics, or equivalently in terms of operators, $S^{\pm}_{\mathbf{r}}=f^{\pm}_{\mathbf{r}}e^{\pm i U_{\mathbf{r}}}$ with $U_{\mathbf{r}}=\sum_{\mathbf{r}^{\prime}\neq\mathbf{r}}\mathrm{arg}(\mathbf{r}-\mathbf{r}^{\prime})f^{\dagger}_{\mathbf{r}}f_{\mathbf{r}}$. The flux attachment for each fermion is enabled by coupling the fermions to a $U(1)$ gauge field described by a CS term~\cite{ruia,ruib}. 
Under the fermionization, the low-energy physics of a frustrated spin system can be derived as the emergent Dirac CS fermions with competing nonlocal interactions induced by gauge field \cite{tsedrak,ruic}. A gapped spin liquid is then formed when certain bosonic orders are generated \cite{ruic}, while the gapless Dirac QSL naturally emerges when the interaction becomes irrelevant \cite{tsedrak,ruic}.

 We specify our study using the Hamilltonian for a 2D XY quantum magnet as starting point,
\begin{equation}\label{eqad1}
H_{0}=\sum_{\mathbf{r},\mathbf{r}^{\prime}}J_{\mathbf{r},\mathbf{r}^{\prime}}(S^x_{\mathbf{r}}S^x_{\mathbf{r}^{\prime}}+S^y_{\mathbf{r}}S^y_{\mathbf{r}^{\prime}}).
\end{equation}
Here, $J_{\mathbf{r},\mathbf{r}^{\prime}}$ includes the first several nearest neighbor (NN) interactions with frustration. After fermionizaition of Eq.\eqref{eqad1}, the CS fermions are cast into the same Hamiltonian as Eq.\eqref{Dirac-fermions}, with additional gauge field-mediated interactions (Fig.1(a)) \cite{ruia,ruib,tsedrakyana,tsedrak}. We now focus on honeycomb lattice, where two Dirac valleys $a=\pm$ emerge at $\mathbf{K}$ and $\mathbf{K}^{\prime}$, related by mirror symmetry, and accordingly $\bm{\tau}^{(+)}=\bm{\tau}$ and $\bm{\tau}^{(-)}=-\bm{\tau}^{T}$ are Pauli matrices defined in the pseudospin (sublattice) space \cite{ruia,ruib,tsedrakyana}, indicating the valley-dependent PSML. For other lattices, there can be more Dirac valleys related by point groups \cite{tsedrak}. We restrict ourselves to studying a stable Dirac QSL \cite{footnote,michael} such that the gauge field-induced interactions between the Dirac fermions are irrelevant operators  \cite{ruia,ruic}.

The CS fermion representation reveals that, the chemical potential $\mu$ of the CS Dirac fermions is tunable by an out-of-plane field $B$. This is because, as long as the Dirac QSL remains stable, the field $B$ generates the out-of-plane polarization that modulates the density of CS fermions $n$ via $\sum_{\mathbf{r}}\langle S^z_{\mathbf{r}}\rangle/N=n-1/2\propto B$ \cite{tsedrakyana,tsedrak}.

Then we consider a quantum impurity, located at $\mathbf{r}_0$ on a lattice bond. This is generally described by fermionic states with spin $\sigma=\uparrow,\downarrow$, orbital $l=1,2,...$, i.e.,
\begin{equation}\label{eqad2}
H_{c}=\sum_{l,\sigma}\epsilon_lc^{\dagger}_{\mathbf{r}_0,l,\sigma}c_{\mathbf{r}_0,l,\sigma}.
\end{equation}
The magnetic interaction between the XY magnet and the impurity is naturally given by 
\begin{equation}\label{eqad3}
H_{int}=\sum_{\mathbf{r}}V(|\mathbf{r}-\mathbf{r}_0|)~ \bm{S}_{\mathbf{r}}\cdot \bm{S}_{c},
\end{equation}
where $V(|\mathbf{r}-\mathbf{r}_0|)$ is nonvanishing only for the near neighbor sites of $\mathbf{r}_0$, and $\bm{S}_{c}=\frac{1}{2}\sum_{l,l^{\prime},\sigma,\sigma^{\prime}}c^{\dagger}_{\mathbf{r}_0,l,\sigma}\boldsymbol{\sigma}_{\sigma,\sigma^{\prime}}c_{\mathbf{r}_0,l^{\prime},\sigma^{\prime}}$ is the impurity spin operator. Here, we assume weak couplings, namely, $V\ll J_{\mathbf{r},\mathbf{r}^{\prime}},|\epsilon_s|$.  The model $H=H_0+H_c+H_{int}$ captures the essential physics including the frustration of magnet and the spin fluctuation of the local impurity.

After tracing out the impurity states, a Heisenberg coupling is generated on the XY magnet between different sublattices. The latter, under the CS fermionization, perturbs the  Dirac spin liquid ground state as an effective Anderson impurity \cite{supment}. The Schrieffer-Wolff transformation then leads to an effective Kondo exchange model as \cite{supment},
\begin{equation}\label{Kondo exchange-v0}
H^{\prime}=\sum_{\mathbf{r}}\lambda(\mathbf{r})f^{\dagger}(\mathbf{r})\frac{\bm{\tau}}{2}f(\mathbf{r})\cdot \mathbf{S}_{imp},
\end{equation}
where we have set $\mathbf{r}_0=0$, $\boldsymbol{\tau}$ denotes the pseudospin (sublattice). $\lambda(\mathbf{r})$ is the coupling strength where a $\mathbf{r}$-dependence is allowed for generality.  $\mathbf{S}_{imp}$ is an effective local spin-1/2 generated by $H_{int}$ \cite{supment}. Interestingly, the CS fermionization translates the original magnetic coupling Eq.\eqref{eqad3} into a Kondo exchange  in the pseudospin  (sublattice) space.

It is natural to assume that $\lambda(\mathbf{r})$ exponentially decays away from the impurity, namely, $\lambda(\mathbf{r})=\lambda_0e^{-|\mathbf{r}|/\xi}$, where $\xi$ is the characteristic scattering length. Then, with projection into the low-energy window, Eq.\eqref{Kondo exchange-v0} takes the form  of Eq.~\eqref{Kondo exchange}. Specifically, the diagonal and off diagonal entries of $\lambda$, $\lambda_d$ and $\lambda_t$, correspond to the intervalley and the intravalley scattering strength, respectively. They are explicitly given by $\lambda_d=2\pi \lambda_0\xi^2$ and $\lambda_t=2\pi \lambda_0\xi^2/(1+|\mathbf{Q}|^2\xi^2)^{3/2}$ with $\mathbf{Q}=\mathbf{K}-\mathbf{K}^{\prime}$ for $\Lambda\ll\xi^{-1}$. Here, $\lambda_t$ is vanishingly small compared to $\lambda_d$  for long-range scattering but is non-negligible for short-range scattering.   

The above shows a systematic mapping from the quantum impurity model in frustrated magnet to the Kondo model in 2D Dirac fermions with valley-dependent PSML, i.e., Eqs.~\eqref{Dirac-fermions} and \eqref{Kondo exchange}.  Accordingly, the reduction to the low-energy soft modes follows, producing  Eq.~\eqref{Kondo_soft-modes} with the diagonal and off-diagonal entries as,  $g_{d/t}=\pi k_F \lambda_{d/t}$.


Before proceeding, we compare the Dirac fermion bath in the spin liquids with that in semimetals \cite{lfritz,andrewk} and the surface states of topological insulators \cite{zitko}. While both have linear dispersion, the Dirac CS fermions in the spin liquids enjoy extraordinary features:  First, the CS fermions are both charge-insulating and spinless. Second, the Kondo exchange acts in the sublattice rather than the true spin space.  Third, the chemical potential is tunable by magnetic field, rather than by the electric potential.

\textit{\color{blue}{Kondo fixed points and thermal conductivity.--}} For the present spin liquid with two Dirac valleys, the pseudospin and valley currents both satisfy $\mathrm{SU}(2)_2$ algebra. The bosonization of the low-energy modes are given by Eq.\eqref{bosonization}. Accordingly, if $g_t$ is negligible, we expect that the impurity is over-screened, and the fermion bath corresponds to the NFL fixed point governed by $\mathrm{U}(1)\times \mathrm{SU}(2)_2\times \mathrm{SU}(2)_2$ CFT. 

Otherwise, with nonvanishing exchange $g_t$, the rotational symmetry in the valley space will be broken. Thus, we introduce $\psi_{1,2}=(\psi_+\pm \psi_{-})/\sqrt{2}$ to diagonalize the Kondo exchange term Eq.\eqref{Kondo_soft-modes} into $H'_{eff}=\sum_{\alpha=1,2}g_\alpha\bm{J}_\alpha(0)\cdot\bm{S}_{imp}$, where $\bm{J}_\alpha=\psi_\alpha^\dagger\bm{\sigma}\psi_\alpha/2$ with $\alpha=1,2$ and $g_{1,2}=g_d\pm d_t$. Accordingly, the two flavors of fermions in Eq.\eqref{Dirac-fermions} should be bosonized individually, which leads to
\begin{equation}
\mathcal{H}_0^{(\alpha)}=\frac{\pi v_F}{2} J^2_\alpha+\frac{2\pi v_F}{3}\bm{J}^2_\alpha.
\end{equation}
The bosonized Hamiltonian $\mathcal{H}_0^{(\alpha)}$ suggests a FL fixed point corresponds to $\mathrm{U}(1)\times \mathrm{SU}(2)_1$ CFT.

The above expectations from CFT can be verified by the perturbative RG calculations. To third order expansion of $g_1$ and $g_2$ \cite{supment}, we obtain the following RG flow, $dg_1/dl=g^2_1-g_1(g^2_1+g^2_2)g_1/2$ and  $dg_2/dl=g^2_2-g_2(g^2_1+g^2_2)g_1/2$. The flow trajectory is shown in Fig.2(a), where two fixed points are revealed as indicated by the red and green dot, respectively. The green dot has one of the couplings been renormalized to zero, thereby describing a  FL fixed point, while the red preserves the symmetric two-channel couplings, suggesting the  NFL behavior.  
\begin{figure}[t]
\includegraphics[width=\linewidth]{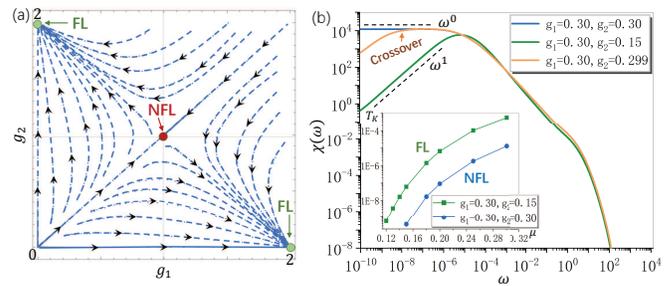}\label{fig2}
\caption{(a) RG flow diagram of the coupling constants $g_1$, $g_2$, which reveal the NFL and FL fixed points. (b) The imaginary part of the impurity dynamical susceptibility of spin calculated by NRG.  The NFL and FL behaviors as well as a crossover are shown in the low temperature regime for different coupling constants. The inset shows the dependence of Kondo temperatures versus chemical potential $\mu$.}
\end{figure}

Using the CFT techniques \cite{Afflecka,Affleckb,Affleckc}, the Green's function (GF) at Kondo fixed points can be calculated by fusion with $\mathbf{S}_{imp}=1/2$ conformal tower. It is obtained that the quasi-particle weight of CS fermions is fully preserved and lost for the FL and NFL fixed point, respectively. The latter predicts an interesting phenomena that the fractionalized excitations in spin liquid lose their quasi-particle nature due to the competing screening channels. 
 Furthermore, via a double fusion procedure \cite{Afflecka,Affleckb,Affleckc}, the scaling behavior of impurity dynamical susceptibility is obtained as  $\chi(\omega)\propto \omega^{0}$ for NFL, while $\chi(\omega)\propto\omega$ for the FL fixed point. 
 
To confirm the CFT results, we set up numerical renormalization group (NRG) calculations starting from the initial Hamiltonian, Eq.\eqref{Dirac-fermions},\eqref{Kondo exchange}. The  calculations are performed using the full-density-matrix NRG~\cite{Andreasa} method implemented in the QSpace tensor library~\cite{Andreasb, Andreasc}. As shown in Fig.2(b), for $g_1\neq g_2$ [$g_1=g_2$], the numerical results of dynamical susceptibility indeed shows $\chi(\omega)\propto\omega$ [$\chi(\omega)\propto \omega^{0}$] at low energies, clearly demonstrating the FL [NFL] fixed point. For $g_1\sim g_2$, a crossover from NFL to FL is  also found. Importantly, the inset of Fig.2(b) shows that the Kondo energy scale $T_K$ is dependent on the chemical potential of CS fermions, implying a tunable Kondo screening by external field $B$. 

Some key features should be pointed out, in contrast with the FL and NFL fixed points in normal metals  \cite{Afflecke}. First, the inherent valleys of Dirac QSLs complicate the situations. The scattering potential among valleys matters. For short-range scattering, the inter-valley scattering is non-negligible, favoring the FL, whereas, the long-range scattering prefers  the NFL fixed point. Second, the CS fermions carry no electron charges and are free from any resistivity anomalies  \cite{Afflecke}.  However, for the FL [NFL] fixed point, the exact screening [overscreening] of the pseudospin takes pace. This should result in an anomalous thermal effect as it generates a many-body local resonance of the CS fermions.

To investigate the Kondo-generated thermal effect, we combine the CFT of CS fermions and the linear response theory \cite{supment,cody}.   
The thermal conductivity can be calculated from the current-current correlation functions,  $\pi(i\omega_n)=-\int^{\beta}_0d\tau e^{i\omega_n\tau}\langle \hat{T}_{\tau}\mathbf{j}_E(\tau)\cdot\mathbf{j}_E(0)\rangle$, where $\mathbf{j}_E(\tau)$ is the thermal current, and the thermal conductance $T\sigma_E(T)=-\lim_{\omega\rightarrow0}\mathrm{Im}\pi^R(\omega)/\omega$, where $\pi^R$ is the retarded correlation function obtained via analytic continuation. Further inserting the self-energy obtained from CFT,  we obtain the thermal conductivity at low temperature as \cite{supment},  
\begin{equation}\label{eq9}
\sigma_E(T)/T=\pi^3\rho_0/[9(1-S)n_{imp}],
\end{equation}  
where we  have assumed a dilute distribution of impurities with density $n_{imp}$, and $S=-1$ [$S=0$] for the FL [NFL] fixed point. Eq.\eqref{eq9} indicates that $\sigma_E(T)$ is sensitive to the DOS of CS fermions at Fermi energy, $\rho_0$,  which is in turn proportional to field $B$, implying a field-modulated thermal conductivity, termed as the magneto-thermal effect.  Moreover, since the phonon's contribution is field-independent, the  predicted Kondo phenomena provides a controllable way to distinguish the intrinsic degrees of freedoms of QSLs. For finite temperature, the higher order corrections from the irrelevant operators in CFT come into play  \cite{Ludwigf,Afflecke}, generating different scaling behaviors for the two different fixed points, i.e.,  $\sigma^{FL}_E(T)/T=\pi^3\rho_0/18n_{imp}-aT^2$ and $\sigma^{NFL}_E(T)/T=\pi^3\rho_0/9n_{imp}-bT^{1/2}$, where $a$, $b$ are universal coefficients.
Therefore, in the crossover regime from NFL to FL shown in Fig.2(b), we expect a non-monotonous thermal conductivity $\sigma_E(T)/T$ versus $T$ when the Kondo resonance is formed.
  
\textit{\color{blue}{Conclusions and discussions.--}} In conclusion, we present a general method, namely a combination of the CS fermionization \cite{ruia,ruib,tsedrakyana} with the WZW theory, to explore novel quantum impurity effects in Dirac QSLs. 
Consequently, FL and NFL behaviors as well as a crossover between them are found,  which can lead to experimental fingerprints for QSLs,
including a Kondo-induced magneto-thermal effect  in the charge-insulating state, a non-monotonous thermal conductivity during the crossover, and an anisotropic spin correlation function because of the PSML. The last one is similar to the pseudospin Kondo-singlet discussed in topological superconductors \cite{ruikondo}. Recent numerical \cite{shijiehu} and analytical studies \cite{tsedrak} suggest that it could be more likely to stabilize the gapless Dirac QSLs in frustrated quantum magnets with a triangular lattice, therefore, it is interesting to search for the predicted Kondo signatures in materials such as $\kappa-\mathrm{ET}_2\mathrm{Cu}_2\mathrm{(CN)}_3$ \cite{shimizu} and $\mathrm{M}[\mathrm{Pd(dmit)_2]_2}$ \cite{myamashitab}.  Finally, we note a recent unusual field-dependence of the muon relaxation at low fields in Zn-brochantite \cite{mogomil}. It might be related to the mechanism discussed here, although further investigations are needed with proper field strength and better materials. 

\begin{acknowledgments}
We thank Andreas Weichselbaum (Brookhaven National Laboratory, USA) and Seung-Sup Lee (Ludwig Maximilian University of Munich, Germany) for providing us the QSpace tensor library and the NRG code for arbitrary type of bath, respectively. Y. W.  was supported by the U.S. Department of Energy, Office of Science, Basic Energy Sciences as a part of the Computational Materials Science Program through the Center for Computational Design of Functional Strongly Correlated Materials and Theoretical Spectroscopy. This work was supported by the Youth Program of National Natural Science Foundation of China (No. 11904225) and  the National Key  R\&D Program of China (Grant No. 2017YFA0303200). 
\end{acknowledgments}


\pagebreak
\vspace{5cm}
\widetext
\setcounter{equation}{0}
\setcounter{figure}{0}
\setcounter{table}{0}
\setcounter{page}{1}
\makeatletter
\renewcommand{\theequation}{S\arabic{equation}}
\renewcommand{\thefigure}{S\arabic{figure}}
\renewcommand{\bibnumfmt}[1]{[S#1]}
\renewcommand{\citenumfont}[1]{S#1}

\pagebreak
\vspace{5cm}
\widetext
\begin{center}
\textbf{\large Supplemental material for: Kondo signatures in Dirac spin liquids: \\
Non-Abelian bosonization after Chern-Simons fermionization}
\end{center}




\date{\today }
\maketitle

\section{The renormaalization effect of local impurity states}
Starting from Eq.(6)-(8) in the main text, we can write down the functional path-integral description of the quantum magnet, and its coupling to the local impurity states. We construct the path integral using the spin coherent state  \cite{efradkan}, which, for $S=1/2$, is denoted by $|n\rangle=\mathrm{exp}[{i\theta}(\hat{z}\times\mathbf{n})\cdot\mathbf{S}]|1/2,1/2\rangle$ and $\langle n| \mathbf{S}|n\rangle=\frac{1}{2}\mathbf{n}$. The zero temperature partition function reads as $Z=\int D\mathbf{n}\int D\overline{c}Dc e^{i(S_0+S^{\prime})}$. Here, $S_0$ is the action describing the frustrated quantum magnet:
\begin{equation}\label{eqsad1}
 S_0[\mathbf{n}]=S\sum_{\mathbf{r}}S_{WZ}[\mathbf{n}(\mathbf{r})]-\int dt\sum_{\langle \mathbf{r},\mathbf{r}^{\prime}\rangle}JS^2\mathbf{n}^{\parallel}(\mathbf{r},t)\cdot\mathbf{n}^{\parallel}(\mathbf{r}^{\prime},t)-...,
\end{equation}
where the $S_{WZ}$ term is the Wess-Zumino term describing the Berry phase accumulation due to the dynamics of each spin operator on the lattice. $\mathbf{n}^{\parallel}=(n^x,n^y)$ because we consider the XY model. Here, the NN term with coupling $J$ is written explicitly and the further neighbor terms are denoted by the ellipsis.  $S^{\prime}$ depicts the local impurity states as well as its local coupling to the nearby spins, which read as,
\begin{equation}\label{eqsad2}
 S^{\prime}=\int dt[\sum_{l,\sigma}\overline{c}_{\mathbf{r}_0,l,\sigma}(i\partial_t-\epsilon_l)c_{\mathbf{r}_0,l,\sigma}-V\sum_{\mathbf{r},l,l^{\prime},\sigma,\sigma^{\prime}}c^{\dagger}_{\mathbf{r}_0,l,\sigma}\boldsymbol{\sigma}_{\sigma,\sigma^{\prime}}c_{\mathbf{r}_0,l^{\prime},\sigma^{\prime}}\cdot\mathbf{n}(\mathbf{r},t)]
 \end{equation}
 where the sum of $\mathbf{r}$ is over the sites nearest to $\mathbf{r}_0$. For simplicity, considering a bipartite lattice with sublattice A and B, and the impurity $\mathbf{r}_0$ being located at the center of the lattice bond, then the sites $\mathbf{r}$ to be summed in the second term can be denoted as $(\mathbf{r}_0,A)$ and $(\mathbf{r}_0,B)$. Now we can integrate out the local quantum dot operators in $S^{\prime}$, which generates the following renormalization to the quantum magnet $S_0$ as,
\begin{equation}\label{eqsad3}
 \Delta S=-i\mathrm{ln}\int D\overline{c}Dce^{i\int^{\infty}_0dt\overline{c}G^{-1}[\mathbf{n}]c},
 \end{equation}
 where we introduced the Grassman field in the basis $c=[c_{\mathbf{r}_0,l,\uparrow},c_{\mathbf{r}_0,l,\downarrow}]^T$ and $\overline{c}=[\overline{c}_{\mathbf{r}_0,l,\uparrow},\overline{c}_{\mathbf{r}_0,l,\downarrow}]$. We defined the inverse of the Green's function (GF), $G^{-1}[\mathbf{n}]=G^{-1}_0-\Sigma[\mathbf{n}]$, where $G_0$, the GF of the local quantum dot, reads in frequency and orbital representation as, $G_{0,l,l^{\prime}}=[\delta_{l,l^{\prime}}/(\omega-\epsilon_l)]\sigma^0$, and $\Sigma_{l,l^{\prime}}[\mathbf{n}]=V(\widetilde{n}^z\sigma^z+\widetilde{n}^-\sigma^++\widetilde{n}^+\sigma^-)$, with $\sigma^{\pm}=\sigma^x\pm i\sigma^y$ and $\widetilde{n}^z=\sum_{\mathbf{r}}n^z(\mathbf{r},\omega)$, $\widetilde{n}^{\pm}=\sum_{\mathbf{r}}n^x(\mathbf{r},\omega)\pm in^y(\mathbf{r},\omega)$. Then, we integrate over the Grassman field $c$, $\overline{c}$, and make expansion the with respect to $\Sigma$, which is proportional to $V$ with $V\ll J,|\epsilon_l|$.  To the leading 
 order of correction (second order), one obtains:
 \begin{equation}\label{eqsad4}
\Delta S=-V^2\int \frac{d\nu}{2\pi}\Pi(\nu)\sum_{\langle \mathbf{r},\mathbf{r}^{\prime}\rangle}\mathbf{n}(\mathbf{r},\nu)\cdot\mathbf{n}(\mathbf{r}^{\prime},-\nu)
  \end{equation}
where the sum of the NN sites is automatically generated.  
$\Pi(\nu)$ is the polarization function due to the fermionic loop of quantum impurity states.  Since we are only interested in the low-energy window in consistent with low-energy CS fermion description of the Dirac QSL, we take the low-frequency approximation of the polarization, which well preserves the low-energy physics for $\nu\ll |\epsilon_l|$. The polarization function is then derived as,
 \begin{equation}\label{eqsadnew4}
\Pi(0)=i\sum_{l,l^{\prime}}\int\frac{d\omega}{2\pi}\frac{1}{(\omega-\epsilon_l e^{-i0^+})(\nu+\omega-\epsilon_{l^{\prime}}e^{-i0^+})}=-\sum_{l,l^{\prime}}\frac{n_F(\epsilon_l)-n_F(\epsilon_{l^{\prime}})}{\epsilon_l-\epsilon_{l^{\prime}}},
  \end{equation}
where $n_F(x)$ is the Fermi-Dirac distribution function. It is obvious that $\Pi(0)>0$ as long as there exists orbitals above and below the Fermi energy. The total renormalized action for the quantum magnet is then obtained as,
\begin{equation}\label{eqsad5}
 S_0[\mathbf{n}]=S\sum_{\mathbf{r}}S_{WZ}[\mathbf{n}(\mathbf{r})]-\int^{\infty}_0dt\sum_{\langle \mathbf{r},\mathbf{r}^{\prime}\rangle}JS^2\mathbf{n}^{\parallel}(\mathbf{r},t)\cdot\mathbf{n}^{\parallel}(\mathbf{r}^{\prime},t)-...-V^2\Pi(0)\int dt\sum_{\langle \mathbf{r},\mathbf{r}^{\prime}\rangle}\mathbf{n}(\mathbf{r},t)\cdot\mathbf{n}(\mathbf{r}^{\prime},t).
\end{equation}
After returning back to the Hamiltonian formalism with respect to the spin operators, the system is then described by:

\begin{equation}\label{eqsad6}
H=\sum_{\mathbf{r},\mathbf{r}^{\prime}}J_{\mathbf{r},\mathbf{r}^{\prime}}(S^x_{\mathbf{r}}S^x_{\mathbf{r}^{\prime}}+S^y_{\mathbf{r}}S^y_{\mathbf{r}^{\prime}})+J^{\prime}\sum_{\langle \mathbf{r},\mathbf{r}^{\prime}\rangle}\mathbf{S}_{\mathbf{r}}\cdot\mathbf{S}_{\mathbf{r}^{\prime}}.
\end{equation}
The quantum impurity states therefore renormalize the frustrated magnets by locally generating a Heisenberg-type exchange coupling of the spin operators with $J^{\prime}=V^2\Pi(0)$. This leads to an additional Ising interaction that is absent in the original XY model. As mentioned, we are considering bipartite lattice with A,B sublattice, thereby the induced Heisenberg term is cast into:
\begin{equation}\label{eqsad7}
H^{\prime\prime}=J^{\prime}\sum_{\langle \mathbf{r},\mathbf{r}^{\prime}\rangle}\mathbf{S}_{\mathbf{r}_0,A}\cdot\mathbf{S}_{\mathbf{r}_0,B}
\end{equation}
where $J^{\prime}\ll J_{\mathbf{r},\mathbf{r}^{\prime}}$ is satisfied since $V\ll  J_{\mathbf{r},\mathbf{r}^{\prime}}$. This term, describing the major effect of the local impurity states, is studied in detail in the remaining part of the supplemental material. In fact, this term is also indicated by the CS fermion description of the Dirac QSLs. In the CS fermion language, the $\mathrm{U}(1)$ gauge field is formed by combination of local string operators defined on neighbor sites \cite{ruia,tsedrakyana,ruib}. The $\mathrm{U}(1)$ gauge field is separated into a non-fluctuating and fluctuating sector \cite{ruia,ruib}, where the latter generates either the long-range spin orderings or the deconfined phases \cite{ruic}.  As required by the CS fermionization method, the non-fluctuating gauge field has to emerge from at least two nearby sites, therefore the local perturbation of the impurity should at least involve two spins on different sublattices, as is shown explicitly by Eq.\eqref{eqsad7} and the above derivations. Noting that the (emergent) sublattice is an indispensable degrees of freedom in the CS fermion description of Dirac spin-liquid, the above derivation is not unique to the studied model but is general, and is applicable when  the quantum impurity states are coupled to a Dirac spin-liquid.

\section{The impurity effect on the Dirac CS fermions}
We assume the formation and stability of a Dirac QSL, so that the first term of Eq.\eqref{eqsad6} is described by deconfined CS fermions as discussed in the main text.   Now we study $H^{\prime\prime}$, for which, we separate explicitly the Ising term and the XY term as,  $H^{\prime\prime}=J^{\prime}_zS^z_{\mathbf{r}_0,A}\cdot S^z_{\mathbf{r}_0,B}+J^{\prime}_{\parallel}(S^x_{\mathbf{r}_0,A}\cdot S^x_{\mathbf{r}_0,B}+S^y_{\mathbf{r}_0,A}\cdot S^y_{\mathbf{r}_0,B})$. Then, it should be noted that the quantum spins at $\mathbf{r}_0$ are also interacting with spins with neighbor sites through $H_0$, from which the relevant term can be extracted  as, $H_{hyb}=\sum_{\mathbf{r}^{\prime}}[J_{\mathbf{r}^{\prime}}(S^x_{\mathbf{r}_0,A}S^x_{\mathbf{r}^{\prime},B}+S^y_{\mathbf{r}_0,A}S^y_{\mathbf{r}^{\prime},B})+J_{\mathbf{r}^{\prime}}(S^x_{\mathbf{r}_0,B}S^x_{\mathbf{r}^{\prime},A}+S^y_{\mathbf{r}_0,B}S^y_{\mathbf{r}^{\prime},A})]$, where $J_{\mathbf{r}^{\prime}}$ accounts for the major contribution from the nearest neighbor exchange coupling. The bath is described by the rest terms, which, in thermal dynamic limit, still stabilizes a Dirac QSL in the long-wave length limit.  Since $J^{\prime}_z$ and $J^{\prime}_{\parallel}$ are much weaker than $J_{\mathbf{r},\mathbf{r}^{\prime}}$, they are treated as local perturbation to the Dirac CS fermions. Then impurity effect is then cast into the Hamiltonian $H^{\prime}=H^{\prime\prime}+H_{hyb}$.  We note that $J^{\prime}_{\parallel}$ terms are weak modifications and can be absorbed into the bath terms, whereas the  Ising interaction $J^{\prime}_z$ term is absent in the XY model.  Then, after representing the quantum spin using CS fermions, $H^{\prime}$ is transformed into:
\begin{equation}\label{eqs1}
  H^{\prime}=J^{\prime}_z(1/2-f^{\dagger}_{\mathbf{r}_0,A}f_{\mathbf{r}_0,A})(1/2-f^{\dagger}_{\mathbf{r}_0,B}f_{\mathbf{r}_0,B})
  +\sum_{\mathbf{r}}J_{\mathbf{r}}(f^{\dagger}_{\mathbf{r}_0,A}e^{iA_{\mathbf{r}_0,\mathbf{r}}}f_{\mathbf{r},B}+f^{\dagger}_{\mathbf{r}_0,B}e^{iA_{\mathbf{r}_0,\mathbf{r}}}f_{\mathbf{r},A}+h.c.),
\end{equation}
where $A_{\mathbf{r}_0,\mathbf{r}}$ is the non-fluctuating lattice gauge field arising from combination of the nearby string operators, which can be taken simply as complex phases and absorbed into $J_{\mathbf{r}}$. let us rename the first and second term in Eq.\eqref{eqs1} as $H_{imp}$ and $H_{hyb}$, respectively, and introduce a different notation for the CS fermions at $\mathbf{r}_0$ as, $c_{\mathbf{r}_0,\overline{\alpha}}=f_{\mathbf{r}_0,\alpha}$, where $\alpha=A,B$ and $\overline{\alpha}$ denotes the opposite sublattice index of $\alpha$. Then, $H^{\prime}$ is cast into,
\begin{eqnarray}
  H_{imp} &=& \sum_{\alpha} \epsilon_f c^{\dagger}_{\mathbf{r}_0,\alpha}c_{\mathbf{r}_0,\alpha}+\frac{U}{2}\sum_{\alpha}c^{\dagger}_{\mathbf{r}_0,\alpha}c_{\mathbf{r}_0,\alpha}c^{\dagger}_{\mathbf{r}_0,\overline{\alpha}}c_{\mathbf{r}_0,\overline{\alpha}}, \\
  H_{hyb} &=& \sum_{\mathbf{r}\alpha}J_{\mathbf{r}}(c^{\dagger}_{\mathbf{r}_0,\alpha}f_{\mathbf{r},\alpha}+h.c.),
\end{eqnarray}
where $\epsilon_f=-U/2=-J^{\prime}_z/2$. Therefore the local perturbation is transformed to a symmetric Anderson impurity with the local effective CS fermions subjected to a Hubbard interaction, which are coupled to the nearby CS fermions of the bath via hopping terms. After a Schrieffer-Wolf transformation, $H^{\prime}$ can be further written in the Kondo regime simply as the exchange coupling to an effective local spin-half impurity $\mathbf{S}_{imp}$ formed by the local CS fermions at $\mathbf{r}_0$, which reads as,
\begin{equation}\label{eqs2}
  H^{\prime}=\sum_{\mathbf{r}}\lambda({\mathbf{r}})f^{\dagger}_{\mathbf{r},\alpha}\boldsymbol{\tau}_{\alpha\beta}f_{\mathbf{r},\beta}\cdot\mathbf{S}_{imp},
\end{equation}
where $\boldsymbol{\tau}$ is the Pauli matrix defined in the pseudospin (sublattice) space. To be more general, we allow $\mathbf{r}$-dependence of the exchange coupling $\lambda({\mathbf{r}})$, describing the scattering potential to the effective quantum impurity. We have shown that the effect of the local quantum states on the frustrated magnet  can be cast into an effective Kondo-exchange in the pseudospin space. This leads to Eq.(9) of the main text. 

\section{mapping to the impurity model coupled to 1D soft fermions}
In the main text, we have shown the general procedure how to map from the impurity model in 2D Dirac fermions to that in 1D soft modes. In this section, we illustrate the detailed transformations using the honeycomb lattice model as the example. 

After projection to long-wave regime near the Dirac CS valleys $\mathbf{K}$ and $\mathbf{K}^{\prime}$, the bath is described as
\begin{equation}\label{eqs3}
  H_0=v_F \sum_{\mathbf{k}}f^{(+)\dagger}_{\mathbf{k},\alpha}(\boldsymbol{\tau}_{\alpha\beta}\cdot\mathbf{k}-\mu)f^{(+)}_{\mathbf{k},\beta}
  +v_F\sum_{\mathbf{k}}f^{(-)\dagger}_{\mathbf{k},\alpha}(-\boldsymbol{\tau}^{\mathrm{T}}_{\alpha\beta}\cdot\mathbf{k}-\mu)f^{(-)}_{\mathbf{k},\beta},
\end{equation}
with valley-dependent PSML. $H^{\prime}$ is projected into the long-wave regime as,
\begin{equation}\label{eqs4}
H^{\prime}_p=g_d\sum_{\mathbf{k},\mathbf{k}^{\prime},a=\pm}f^{(a)\dagger}_{\mathbf{k},\alpha}\boldsymbol{\tau}_{\alpha\beta}\cdot \mathbf{S}_{imp},f^{(a)}_{\mathbf{k}^{\prime},\beta}
+g_t\sum_{\mathbf{k},\mathbf{k}^{\prime},a=\pm}f^{(a)\dagger}_{\mathbf{k},\alpha}\boldsymbol{\tau}_{\alpha\beta}\cdot \mathbf{S}_{imp},f^{(\overline{a})}_{\mathbf{k}^{\prime},\beta}.
\end{equation}
where $a$ denotes the two valleys. One firstly make a unitary transformation to diagonalize CS fermions at each Dirac valley, $H^{\prime}_p$ transform accordingly under the unitary transformation, leading to
\begin{equation}\label{eqs4}
 H^{\prime}_p=g_d\sum_{\mathbf{k},\mathbf{k}^{\prime},a=\pm} c^{(a)\dagger}_{\mathbf{k}} U^{(a)}(\theta_{\mathbf{k}})\tau^iU^{(a)\dagger}(\theta_{\mathbf{k}^{\prime}}) c^{(a)}_{\mathbf{k}^{\prime}}S^i_{imp}
+g_t\sum_{\mathbf{k},\mathbf{k}^{\prime},a=\pm} c^{(a)\dagger}_{\mathbf{k}} U^{(a)}(\theta_{\mathbf{k}})\tau^iU^{(\overline{a})\dagger}(\theta_{\mathbf{k}^{\prime}}) c^{(\overline{a})}_{\mathbf{k}^{\prime}}S^i_{imp},
\end{equation}
where $c^{(a)}_{\mathbf{k}}$ is the transformed spinor in band (sublattice) space at valley $a$, $U^{(a)}(\theta_{\mathbf{k}})$ the unitary rotation matrix applied for fermions at valley $a$ which is only dependent on the angle of momentum $\theta_{\mathbf{k}}$. Then, utilizing the rotational symmetry of the impurity scattering, we transform the fermions to the orbital angular momentum partial waves using $c^{(a)}_{\mathbf{k}}=\sum_l e^{il\theta}c^{(a)}_{l,k}/\sqrt{2\pi k}$, where $l$ is the partial wave index. After insertion of the specific form of the unitary rotation matrix $U^{(a)}(\theta_\mathbf{k})$, the integral over the polar angle automatically picks up several different partial waves $l$, generating the following coupling as,
\begin{equation}\label{eqs5}
 H^{\prime}_p=g_d\sum_{a} \int dkdk^{\prime}\sqrt{kk^{\prime}}
 c^{(a)\dagger}_{l,k} U^{(a)}(l)\boldsymbol{\tau}\cdot \mathbf{S}_{imp} U^{(a)\dagger}_l c^{(a)}_{l,k^{\prime}}
+g_t\sum_{a} \int dkdk^{\prime}\sqrt{kk^{\prime}}
 c^{(a)\dagger}_{l,k} U^{(a)}(l)\boldsymbol{\tau}\cdot \mathbf{S}_{imp} U^{(\overline{a})\dagger}_l c^{(\overline{a})}_{l,k^{\prime}},
\end{equation}
where constants have been absorbed into the tuning parameter $g_d$ and $g_t$, the sum over repeated notations such as $l$ is implicit. $U^{(a)}_l$ is the rotation matrix again transformed to the angular orbital momentum space, whose components are delta functions that select the channel $l$ relevant to the impurity, i.e.,
\begin{equation}\label{eqs6}
U^{(a)}_l=\frac{1}{\sqrt{2}}\left(
\begin{array}{cc}
\delta_{l,0} & a\delta_{l,-a}\\
\delta_{l,0} & -a\delta_{l,-a}\\
\end{array}
\right),
\end{equation}
where $a=\pm$ denotes the two valleys. Eq.\eqref{eqs5} implies that the impurity is coupled to an effective CS fermions $d^{(a)}_{\mathbf{k}}=\sum_l U^{(a)\dagger}(\theta_{\mathbf{k}})c^{(a)}_{l,k}$, which is combinations of  1D CS fermions with different index $l$ for different valleys. $l=0,-1$ are coupled to the impurity at $\mathbf{K}$ whle $l=0,1$ are involved at $\mathbf{K}^{\prime}$ valley. Therefore, the impurity only picks up these relevant $l$ channels. Since the bath, after rotation to the angular orbital momentum space, enjoy independent $l$ components with $l$ being good quantum number due to the rotational invariance of the problem, we can select from the bath these relevant channels, leading to,
\begin{equation}\label{eqs7}
H_0=\sum_{l=-1,0}\int^{\infty}_0 dk(\epsilon_{k,\alpha}-\mu)c^{(+)\dagger}_{k,l,\alpha}c^{(+)}_{k,l,\alpha}+\sum_{l=0,1}\int^{\infty}_0 dk(\epsilon_{k,\alpha}-\mu)c^{(-)\dagger}_{k,l,\alpha}c^{(-)}_{k,l,\alpha},
\end{equation}
where $\epsilon_{k,\alpha}=\alpha v_Fk$. It is convenient to introduce the energy representation for the impurity problem \cite{zitko}, and define the effective CS fermions with combination of operators for the conduction and valence Dirac band as,
\begin{eqnarray}
  d^{(+)}_{\epsilon} &=& \frac{1}{\sqrt{2}}[c^{(+)}_{\epsilon,0,+}\theta(\epsilon)+c^{(+)}_{\epsilon,0,-}\theta(-\epsilon), c^{(+)}_{\epsilon,-1,+}\theta(\epsilon)-c^{(+)}_{\epsilon,-1,-}\theta(-\epsilon)]^{\mathrm{T}}, \\
  d^{(-)}_{\epsilon}&=& \frac{1}{\sqrt{2}}[c^{(-)}_{\epsilon,0,+}\theta(\epsilon)+c^{(-)}_{\epsilon,0,-}\theta(-\epsilon), -c^{(-)}_{\epsilon,1,+}\theta(\epsilon)+c^{(-)}_{\epsilon,1,-}\theta(-\epsilon)]^{\mathrm{T}}.
\end{eqnarray}
Using $d^{(a)}_{\epsilon}$, $H_0$ is cast into a simple form as,
\begin{equation}\label{eqs8}
H_0=\sum_{a,\sigma}\int^{\infty}_{-\infty} d\epsilon(\epsilon-\mu)d^{(a)\dagger}_{\epsilon,\sigma}d^{(a)}_{\epsilon,\sigma},
\end{equation}
where $v_F$ is set to 1. $d^{(a)}_{\epsilon,\sigma=1,2}$ are the two entries of the spinor defined in Eq.(19) and (20). Accordingly, the hybridization term $H^{\prime}_p$ is reduced to the following form as,
\begin{equation}\label{eqs9}
 H^{\prime}_p=g_d\sum_{a} \int^{+\infty}_{-\infty} d\epsilon d\epsilon^{\prime} [\rho(\epsilon)\rho(\epsilon^{\prime})]^{1/2}
 d^{(a)\dagger}_{\epsilon} \boldsymbol{\tau}\cdot \mathbf{S}_{imp} d^{(a)}_{\epsilon^{\prime}}
+g_t\sum_{a} \int^{+\infty}_{-\infty} d\epsilon d\epsilon^{\prime} [\rho(\epsilon)\rho(\epsilon^{\prime})]^{1/2}
 d^{(a)\dagger}_{\epsilon} \boldsymbol{\tau}\cdot \mathbf{S}_{imp} d^{(\overline{a})}_{\epsilon^{\prime}},
\end{equation}
where $\rho(\epsilon)=|\epsilon|/2\pi v^2_F$ is the density of states of Dirac CS fermions, leading to a pseudogap in the above hybridizations. Detailed studies on the pseudogapped cases have shown that the strong coupling fixed points at zero temperature are not modified by approximating the density of states by that of the Fermi energy \cite{Ruid}, as long as $\mu\neq0$. With this approximation, one can absorb the density of states into the couplings and rename the fermionic field as $\psi^a$. This leads to an Kondo-exchange model coupled to 1D chiral soft modes, in consistent with the general form, i.e., Eq.(3),(4) in the main text.

\section{derivation of the decoupled Wess-Zumino-Witten CFT using non-Abelian gauge invariance}
The infrared fixed point of the reduced model (impurity coupled to the 1D soft modes) is described by a Wess-Zumino-Witten (WZW) CFT. We now show in this section that the underlying CFT has a decoupled multichannel structure and can be derived simply from the principle of gauge invariance.
The following contents are separated into three steps including the derivation of Ward identities from the non-Abelian gauge symmetry, the chiral symmetry, and the deduction of the exact functional free energy.
\subsection{Ward Identity from non-abelian gauge transformations}

From the mapped 1D model of soft modes, we can start from a Dirac field in a representation $r$ of a Lie group $G$ coupled with a given gauge field $A$,namely $\mathcal{L}=\bar{\psi} (i\slashed D-m)\psi$ with $D_\mu=\partial_\mu-igA_\mu$.
We define the free energy $W$ as
\begin{equation}
e^{-iW[A]}=Z[A]=\int  \mathcal{D}\psi\mathcal{D}\bar{\psi} ~e^{iS}.
\end{equation}
The classical theory is invariant under the gauge transformations, $\psi \rightarrow U\psi$,$\bar{\psi} \rightarrow \bar{\psi}U^{-1}$, and $A_\mu \rightarrow A^{U}=U A_\mu U^{-1}+\frac{i}{g} U\partial_\mu U^{-1}$, whose infinitesimal version is
$ \psi \rightarrow (1+i\alpha)\psi$, $\bar{\psi} \rightarrow \bar{\psi} (1-i\alpha)$, and
$A_\mu \rightarrow A_\mu +\frac{1}{g}\mathcal{D}_\mu \alpha$. Assuming that the functional measurement $\mathcal{D}\bar{\psi} \mathcal{D}\psi$ is also gauge invariant, the free energy satisfies
\begin{equation}
W[A]=W[A^U]+2\pi n[U]
\end{equation}
with $n$ being an integer determined by $U$, which vanishes for infinitesimal transformations. Accordingly,
\begin{equation}
\begin{split}
0 &= W[A_\mu +\mathcal{D}_\mu \alpha]-W[A_\mu]= \int dx \frac{\delta W}{\delta A_\mu^a}(\mathcal{D}_\mu \alpha)^a= -\int dx ~ \mathrm{tr}\alpha\mathcal{D}_\mu\frac{\delta W}{\delta A_\mu},
\end{split}
\end{equation}
which implies
\begin{equation}
\mathcal{D}_\mu\frac{\delta W}{\delta A_\mu}=0.
\end{equation}
The variation of the free energy to the gauge field is then calculated as
\begin{equation}
\begin{split}
-i\frac{\delta W}{\delta A_\mu}&=\frac{1}{Z}\frac{\delta Z}{\delta A_\mu}=\frac{1}{Z} \int \mathcal{D}\psi\mathcal{D}\bar{\psi} ~ i\frac{\delta S}{\delta A_\mu}e^{iS}= i\frac{1}{Z} \int \mathcal{D}\psi\mathcal{D}\bar{\psi} ~ J^\mu e^{iS}= i\langle J^\mu \rangle_{A}.
\end{split}
\end{equation}
Thus we prove the Ward identity
\begin{equation}
\mathcal{D}_{\mu}\langle J^\mu \rangle=0.
\end{equation}

\subsection{Ward Identity from chiral invariance}
Now we study the chiral gauge transformations given by $\psi \rightarrow \psi'=(1+i\alpha \gamma^5)\psi$m
$\bar{\psi} \rightarrow \bar{\psi}'=\bar{\psi}(1+i\alpha \gamma^5)$ and
$A_\mu \rightarrow A_\mu^\prime =A_\mu + \mathcal{D}_\mu \alpha\gamma^5$.
It is straightforward to check that the classical theory is invariant under the gauge transformations. However the functional measurement does not respect the transformations, leading to a Jacobian determinant $\mathcal{J}$. Thus the Ward identity should be modified because of $W[A^\prime]\ne W[A]$. From the partition function, we obtain,
\begin{equation}
\begin{split}
Z[A]&=\int \mathcal{D}\psi\mathcal{D}\bar{\psi}~ e^{iS[\psi,\bar{\psi},A]}=\int \mathcal{D}\psi'\mathcal{D}\bar{\psi'}~ \mathcal{J}~ e^{iS[\psi',\bar{\psi}',A']}= Z[A']+\int \mathcal{D}\psi\mathcal{D}\bar{\psi}\int dx~\frac{\delta \mathcal{J}}{\delta \alpha^a}|_{\alpha(x)=0}\alpha^a(x)e^{iS[\psi,\bar{\psi},A']},
\end{split}
\end{equation}
leading to
\begin{eqnarray}
\frac{Z[A']-Z[A]}{Z[A]}=-\langle \int dx~\frac{\delta \mathcal{J}}{\delta \alpha^a}|_{\alpha(x)=0}\alpha^a(x)\rangle.
\end{eqnarray}
Besides, we have 
\begin{eqnarray}
\frac{Z[A']-Z[A]}{Z[A]}=-ig \langle \int dx (\mathcal{D}_\mu J^{5\mu})^a \alpha^a\rangle.
\end{eqnarray}
Thus, conservation equation for the axial current is obtained as,
\begin{equation}
\mathcal{D}_\mu J^{5\mu}=-i\frac{1}{g} \frac{\delta\mathcal{J}}{\delta\alpha}|_{\alpha=0}.
\end{equation}
The remaining task is then to evaluate the Jacobian determinant.  This can be readily done using the method developed by Fujikawa, which is also utilized in a similar situation of 3+1D with the chiral anomaly. A straightforward calculation in 1+1D then generates the Ward identity for the axial current as,
\begin{equation}
\mathcal{D}_\mu J^{5\mu}=-\frac{C(r)}{2\pi}\epsilon^{\mu\nu}F_{\mu\nu}.
\end{equation}
where in the derivation we have defined the Dirac matrices $\gamma^0=\sigma^2$, $\gamma^1=i\sigma^1$ and $\gamma^3=\gamma^0\gamma^1=\sigma^3$ and used  $\mathrm{tr}(t^at^b)=C(r)\delta^{ab}$ with $C(r)$ a constant for eacg representation $r$ with $t^a$ the representation matrix.

\subsection{The exact functional determinant in two dimensions}
Noting  that there exists a unique relation only in 1+1D dimensions,
$\gamma^\mu \gamma^3=-\epsilon^{\mu\nu}\gamma_\nu$, which enables us to rewrite the chiral current as
$J^{3\mu}=-\epsilon^{\mu\nu}J_\nu$. 
Therefore, the two Ward identities derived above are collected into a united form of the CS fermion current as,
\begin{eqnarray}
\mathcal{D}_\mu J^\mu&=& 0\\
\epsilon^{\mu\nu} \mathcal{D}_\mu J_\nu &=& \frac{C(r)}{2\pi}\epsilon^{\mu\nu}F_{\mu\nu}.
\end{eqnarray}
Now the uniqueness of dimension two, compared with higher dimensions, lies in that the current $J^\mu$ is completely determined by the two Ward identities. Before solving the equations we first introduce the chiral coordinates,
$x^{+}=x^0+x^1$, $x^{-}=x^0-x^1$.
In the chiral coordinates the metric $\eta$ and total anti-symmetric tensor $\epsilon$ are represented, respectively, as
\begin{equation}
\eta_{\mu\nu}=\begin{pmatrix}
0 & \frac{1}{2}\\
\frac{1}{2} & 0
\end{pmatrix}, \quad \epsilon^{\mu\nu}=\begin{pmatrix}
0 & 2\\
-2 & 0
\end{pmatrix}.
\end{equation}
Accordingly we define $J^+=J^0+J^1$, $J^-=J^0-J^1$,  and $A^+=A^0+A^1$, $ A^-=A^0-A^1$. In these notations the two identities can be cast into the following symmetric form,
\begin{eqnarray}
\partial_{+} J_{-}-i[A_{+},J_{-}]&=&\frac{C(r)}{2\pi}F_{+-}\\
\partial_{-} J_{+}-i[A_{-},J_{+}]&=&\frac{C(r)}{2\pi}F_{-+}.
\end{eqnarray}
To obtain the explicit form of the solution, we introduce the expression for the gauge fields,
$A_{+}=ig^{-1}\partial_{+}g$, $A_{-}=ih^{-1}\partial_{-}h$.
with $g$ and $h$ being fields of group elements in $G$. Then it is straightforward to check that
\begin{eqnarray}
J_{+}&=&\frac{C(r)}{2\pi}(ig^{-1}\partial_{+}g-ih^{-1}\partial_{+}h)\\
J_{-}&=&\frac{C(r)}{2\pi}(ih^{-1}\partial_{-}h-ig^{-1}\partial_{-}g),
\end{eqnarray}
are the solutions of the equations.

As promised we shall work out an explicit expression of the free energy $W[A]$, which is gauge invariant, using the fields $g$ and $h$. The field $t(x)\in G$ gives the gauge transformations,
\begin{eqnarray}
A_{+}=ig^{-1}\partial_{+}g &\longrightarrow& itg^{-1}\partial_{+}gt^{-1}+it\partial_{+}t^{-1}=i(gt^{-1})^{-1}\partial_{+}(gt^{-1})\\
A_{-}=ih^{-1}\partial_{-}h &\longrightarrow& ith^{-1}\partial_{-}ht^{-1}+it\partial_{-}t^{-1}=i(ht^{-1})^{-1}\partial_{-}(ht^{-1}),
\end{eqnarray}
which are translated to $g$ and $h$ as
\begin{equation}
(g,h)\longrightarrow (g,h)t^{-1}.
\end{equation}
The gauge invariance of $W[A]$ is now expressed as
\begin{equation}
W[g,h]=W[gt^{-1},ht^{-1}],
\end{equation}
for any field $t(x)$. So it is sufficient to work with the gauge $A_{-}=0$, or equivalently $h$ constant.
\begin{equation}
\delta W=-\frac{1}{\pi}\int dx~\mathrm{tr}(g^{-1}\partial_{-}g~\delta (g^{-1}\partial_{+}g))=-\frac{1}{\pi}\int dx~ \left(\mathrm{tr}(\partial_{+}\partial_{-}g^{-1}\delta g)-\mathrm{tr}(g^{-1}\partial_{+}g~g^{-1}\partial_{-}g~g^{-1}\delta g)\right)
\end{equation}
Noting that
\begin{equation}
\begin{split}
\delta \int dx~\mathrm{tr\partial_{-}g^{-1}\partial_{+}g}&=-2\int dx~ \mathrm{tr}(\partial_{+}\partial_{-}g^{-1})\delta g + \int dx~ \mathrm{tr}(g^{-1}\partial_{+}g~g^{-1}\partial_{-}g~g^{-1}\delta g)\nonumber\\
&+\int dx~ \mathrm{tr}(g^{-1}\partial_{-}g~g^{-1}\partial_{+}g~g^{-1}\delta g),
\end{split}
\end{equation}
 we have
\begin{equation}
\delta W=-\frac{1}{8\pi}\delta\int dx~ \mathrm{tr}(g^{-1}\partial^\mu g~ g^{-1}\partial_\mu g)+\frac{1}{4\pi}\int dx~\epsilon^{\mu\nu}\mathrm{tr}(g^{-1}\partial_{\mu}g~g^{-1}\partial_{\nu}g~g^{-1}\delta g).
\end{equation}
Let us assume that $G=SU(N)$ and the spacetime manifold is compactified as $S^2$. Then it is well-known that the second term on the right hand of the above equation is a variation of a Wess-Zumino term. Thus the free energy can be explicitly written as
\begin{equation}
W[g]=-\frac{1}{8\pi}\int d^2x~ \mathrm{tr}(g^{-1}\partial^\mu g~ g^{-1}\partial_\mu g)+\frac{1}{12\pi}\int d\tau d^2x~\epsilon^{\mu\nu\rho}\mathrm{tr}(\tilde{g}^{-1}\partial_{\mu}\tilde{g}~\tilde{g}^{-1}\partial_{\nu}\tilde{g}~\tilde{g}^{-1}\partial_\rho \tilde{g}),
\end{equation}
where $\tilde{g}(\tau,x)$ with $\tau\in[0,1]$ is a continuous extension of $g(x)$ with $\tilde{g}(0,x)=g(x)$ and $\tilde{g}(1,x)$ being constant.

Last, for the Dirac fields with both the spin and flavor as in Eq.(7) of the main text, the above derivation works but needs to be generalized with coupling to two non-Abelian gauge field $A_{\mu}$ and $B_{mu}$, resulting in the following Langrangian as,
\begin{equation}
\begin{split}
\mathcal{L} &=\bar{\psi}^{ia} (i\gamma^\mu \partial_\mu\delta^{ij}\delta^{ab}+A_\mu^{ij}\delta^{ab}+\delta^{ij}B_\mu^{ab})\psi^{jb}
   =\bar{\psi}(i\gamma^\mu \partial_\mu 1_n\otimes 1_m+A_\mu\otimes 1_m+1_n\otimes B_\mu)\psi,
\end{split}
\end{equation}
where accordingly we have$ A_+=ig_A^{-1}\partial_{+} g_A$, $A_-=ih_A\partial_{-} h_A$,
$B_+=ig_B^{-1}\partial_{+} g_B$,  and $B_-=ih_B\partial_{-} h_B$, such that
\begin{eqnarray}
A_+\otimes 1_m+1_n\otimes B_+=ig_A^{-1}\otimes g_B^{-1}\partial_+(g_A\otimes g_B)\\
A_-\otimes 1_m+1_n\otimes B_-=ih_A^{-1}\otimes h_B^{-1}\partial_+(h_A\otimes h_B)
\end{eqnarray}
For two arbitrary matrices $M$ and $N$, one has the following property $\mathrm{tr}(M\otimes N)=\mathrm{tr}(M)\mathrm{Tr}(N)$. 
Moreover, for $g\in SU(n)$, $\mathrm{tr}(g\partial_\mu g^{-1})=0$ since the Lie algebra consists of $n\times n$ traceless Hermitian matrices. With the above two identities, it is straightforward to derive that the following WZW emerges:
\begin{equation}
W[g_A \otimes g_B]=MW[g_A]+NW[g_B].
\end{equation}
This is the decoupled WZW CFT, from which one can read of the fusion rules in order to obtain the Kondo fixed points, which are discussed in the main text for both cases, i.e., with and without the off-diagonal entries of the scattering $g_{ab}$ (Eq.(4) of the main text).

\section{perturbative RG calculation of $\beta$-functions of exchange couplings}
In order to determine the fixed points at the strong-coupling regime, we perform a perturbative RG calculation of the $\beta$-functions with respect to the derived effective 1D model, which reads as $H=H_0+H^{\prime}_p$, where $H_0$ is the rotated Hamiltonian in the valley space with respect to Eq.\eqref{eqs7}, which is of the form,
\begin{equation}\label{eqsn10}
H_0=\sum_{m=1,2}\sum_{\sigma}\int^{\infty}_{-\infty} d\epsilon(\epsilon-\mu)d^{\dagger}_{\epsilon,m,\sigma}d_{\epsilon,m,\sigma},
\end{equation}
and $H^{\prime}_p$ is approximated by using the density of states at the Fermi energy, $\rho_0$, leading to
\begin{equation}\label{eqsn11}
 H^{\prime}_p=g_1\int^{\infty}_{-\infty}d\epsilon d\epsilon^{\prime}d^{\dagger}_{\epsilon,1,\sigma}\boldsymbol{\tau}_{\sigma\sigma^{\prime}}\cdot\mathbf{S}_{imp}d_{\epsilon^{\prime},1,\sigma^{\prime}}
 +g_2\int^{\infty}_{-\infty}d\epsilon d\epsilon^{\prime}d^{\dagger}_{\epsilon,2,\sigma}\boldsymbol{\tau}_{\sigma\sigma^{\prime}}\cdot\mathbf{S}_{imp}d_{\epsilon^{\prime},2,\sigma^{\prime}},
\end{equation}
where a rotation in the valley space is performed, leading to the channel $m=1,2$, and $g_1=\rho_0(g_d+g_t)$, $g_2=\rho_0(g_d-g_t)$. The perturbative expansion over the two terms in $H^{\prime}_p$ can be constructed with Feynman diagrams to the two-loop order. Integrating out the fast mode momentum leads to the renormalization group flow as,
\begin{eqnarray}
  dg_1/dl &=& g^2_1-g_1(g^2_1+g^2_2)g_1/2, \\
  dg_2/dl&=&g^2_2-g_2(g^2_1+g^2_2)g_1/2.
\end{eqnarray}
where $dl=d\Lambda/\Lambda$ is the RG scaling parameter. The first term obtained from second order is relevant, showing the asymptotic free of the exchange coupling, and the second term from the third order contributes a suppression of the relevant flow, generating a channel-mixed fixed point with finite values of $g$'s, as shown by Fig.2(a) of the main text.

\section{The thermal conductivity from CS fermions}
We now list in this section details for calculation of the thermal conductivity, with respect to both of the two Kondo-generated fixed points. The starting point is Eq.\eqref{eqsn10} and $\eqref{eqsn11}$, which have been bosonized in the non-Abelian fashion before. The 1D model can be further mapped to half-infinite chain with left and right movers denoted by fields $d_{L/R,m,\sigma}$ \cite{Afflecka}. We consider the single-particle Green's function defined by $\langle d^{\dagger}_{L,m,\sigma}(z_1)d_{R,m,\sigma}(z_2)\rangle$, where $z$ lies in the complex plane representing for the 1+1D spacetime. We obtain $\langle d^{\dagger}_{L,m,\sigma}(z_1)d_{R,m,\sigma}(z_2)\rangle=0$ for the case with no boundary, and $\langle d^{\dagger}_{L,m,\sigma}(z_1)d_{R,m,\sigma}(z_2)\rangle_{Free}=1/(z_1-\overline{z}_2)$ for a trivial boundary (corresponding to the weak coupling regime). The correlation with respect to the Kondo fixed point can be calculated via the boundary state that in turn obtained by fusion \cite{Afflecka,Affleckb,Affleckc,Ludwig,Affleckd,Afflecke,Ludwigf}, leading to the scattering matrix $S$ connecting the correlations as $\langle d^{\dagger}_{L}(z_1)d_R(z_2)\rangle_{Kondo}=S\langle d^{\dagger}_{L,m,\sigma}(z_1)d_{R,m,\sigma}(z_2)\rangle_{Free}$ , where  $S=-1$ and $S=0$ for the FL and NFL fixed point.
Assuming a dilute impurity with density $n_{imp}$, the scattering time $\tau^{-1}_s=-2\mathrm{Im}\Sigma^R(\omega)$, where $\Sigma^R(\omega)$ is the retarded self-energy that is related to $S$ \cite{Afflecke}.

On the other hand, the thermal current can be readily derived from Eq.\eqref{eqsn10} and Eq.\eqref{eqsn11} as,
\begin{equation}\label{eqsn12}
\mathbf{j}_E=-v_F\hat{k}\sum_{m,\sigma}\int d\epsilon (\epsilon-\mu)d^{\dagger}_{\epsilon,m\sigma}d_{\epsilon,m,\sigma},
\end{equation}
where $\hat{k}=\mathbf{k}/k$ and $v_F$ is set to 1 in the following. The thermal conductivity can be evaluated through $T\sigma_E(\omega)=-\lim_{\omega\rightarrow0}\mathrm{Im}\pi^R(\omega)/\omega$, where $\pi^R$ is the retarded current-current correlation. We firstly calculate the current correlation function in Matsubara form as,
\begin{equation}\label{eqsn13}
\pi(i\omega_n)=-\frac{1}{3}\int^{\beta}_0 d\tau e^{i\omega_n\tau} \langle \hat{T}_{\tau}\mathbf{j}_E(\tau)\cdot\mathbf{j}_E(0)\rangle,
\end{equation}
where $\hat{T}$ is the imaginary time ordering. Inserting Eq.\eqref{eqsn12} into Eq.\eqref{eqsn13}, we obtain
\begin{equation}\label{eqsn14}
\pi(i\omega_n)=\frac{1}{3}\int d\epsilon\xi^2\frac{1}{\beta}\sum_{i\nu_n} g(\epsilon,i\omega_n+i\nu_n)g(\epsilon,i\nu_n),
\end{equation}
where $\xi=\epsilon-\mu$, and $g(\epsilon,i\omega_n)$ is the Matusbara Green's function of the d-fermions.  Performing the sum of Matusbara frequency $i\nu_n$ and after insertion of $\pi(i\omega_n)$ into $\sigma_E$, one obtains that
\begin{equation}\label{eqsn15}
T\sigma_E=\frac{1}{3}\int d\epsilon\xi^2\int d\nu\delta(\epsilon-\nu)\tau_s(-\frac{\partial}{\partial{\nu}}n_F(\nu)),
\end{equation}
where $n_F(\nu)=1/\mathrm{exp}[\beta(\nu-\mu)+1]$ is the Fermi distribution function.  We consider lowest order $T$ behavior of $T\sigma_E$. Since $\tau_s$ is dependent on $S$ which further relies on higher order $T$-terms, its $T$-dependence can be neglected for low $T$. After inserting the Fermi distribution function and completing the integrals, it is straightforward to find that the right-hand-side of Eq.\eqref{eqsn15} is proportional to $T^2$, leading to $\sigma_E(T)/T=\pi^2\tau_s/9=(\pi^3\rho_0)/[9(1-S)n_{imp}]$ for low $T$. With further taking into account the higher order $T$-dependence in $S$ \cite{Ludwigf,Afflecke}, one can obtain the different $T$-scalings of the thermal conductivity for the two fixed points, as shown by the main text.

\section{Details of numerical renormalization group calculations}
For the honeycomb lattice XY model discussed above, we derived an effective two-channel Kondo model with impurity spin $\mathbf{S}_{\text{imp}}=\frac{1}{2}$ and Kondo couplings $g_1$ and $g_2$.  This can be solved using numerical renormalization group. The density of states of the non-interacting bath takes a linear form,
\begin{equation}
\rho(\epsilon)=\frac{|\epsilon-\mu|}{2\pi},
\end{equation}
where, $\epsilon\in[-1, 1]$ and $\mu$ is the chemical potential. The bath is discretized logarithmically and mapped to a semi-infinite ``Wilson chain" with exponentially decaying hoppings, and the impurity coupled to the first chain site via Kondo constants $g_1$ and $g_2$. The chain is diagonalized iteratively while discarding high-energy states, thereby zooming in on low-energy properties: the finite-size level spacing of a chain ending at site $k$ is of order $\omega_k\propto \Lambda^{-k/2}$. Here $\Lambda>1$ is a discretization parameter, chosen to be 2 in this work. We use the full-density-matrix NRG~\cite{Andreasa} method to solve this model, exploiting its full U(1)$_{\text{charge}}$ $\times$ SU(2)$_{\text{spin}}$ $\times$ SU(2)$_{\text{channel}}$ symmetry when $g_1=g_2$ and U(1)$_{\text{charge}}$ $\times$ SU(2)$_{\text{spin}}$ when $g_1\ne g_2$ using the QSpace~\cite{Andreasb,Andreasc} tensor library.  We keep 4000 multiplets in the diagonalizations. The imaginary part of the impurity dynamical susceptibility, $\chi(\omega)=-\text{Im}\langle S||S\rangle_{\omega}$, was calculated at temperature $T=10^{-10}$. The FL [NFL] Kondo scale is determined as the energy at which $\chi(\omega)$ [$\frac{\text{d} (\log\chi(\omega))}{\text{d}(\log\omega)}$] has a maximum.



\begin{thebibliography}{99}
\bibitem{footnote2} The Dirac valleys can also be related by projected symmetry group in the spin liquid cases.
\bibitem{footnote} The stability of Dirac QSLs under the gauge fluctuation has been studied and justified in the large-N case \cite{michael}. The Dirac QSL cannot be stabilized as ground state on honeycomb lattice with two valleys (but may emerge at the critical point of the unconventional phase transition \cite{ruic}). However, since our aim is on the impurity effect, we loosen the stability condition and assume that an impurity is coupled to the Dirac QSL state with two Dirac valleys. This paradigmatic example can clearly demonstrate the method and results that are applicable for general QSLs with more Dirac valleys.
\end{thebibliography}
\end{document}